\documentclass[letterpaper, 10 pt, conference]{ieeeconf}  
\usepackage{amsmath}
\usepackage{graphicx}
\IEEEoverridecommandlockouts                              

\usepackage{cite}
\usepackage{color}
\usepackage{soul}
\usepackage[hidelinks]{hyperref}
\usepackage{color}
\title{\LARGE \bf
Task aware Dynamic Movement Primitives for failure detection and recovery in contact rich manipulation
}

\author{Bhavnashri A$^{1}$ Sobia Shafi$^{2}$ Krishnapuram Himavarshini $^{3}$ and Anuj Tiwari$^{2}$
\thanks{This work has been submitted to the IEEE for possible publication.
Copyright may be transferred without notice, after which this version may
no longer be accessible.}
\thanks{\noindent
$^{1}$Atomberg Technologies Limited, India
        {\tt\small bhavnashri.a@gmail.com}}%
\thanks{\noindent $^{2}$Indian Institute of Technology Madras (IITM), India
        {\tt\small id25d018@smail.iitm.ac.in,anujt@iitm.ac.in}}%
\thanks{\noindent $^{3}$Appian Corporation, India
        {\tt\small himavarshini2704@gmail.com}}%
 }

\begin{document}

\maketitle
\thispagestyle{empty}
\pagestyle{empty}

\begin{abstract}
Assembly remains a challenging robotic manipulation task in presence of tight tolerances and complex contact interactions. While Learning from Demonstration (LfD) frameworks like Dynamic Movement Primitives (DMPs) can effectively encode trajectories from a single demonstration, they are highly sensitive to variations in initial grasp configurations and external contact forces. Such variations often lead to task failures during the contact rich phases. This paper presents a task aware failure detection and recovery framework that integrates DMP based trajectory generation with real time stage classification. Utilizing Quadratic Discriminant Analysis (QDA) trained on multimodal sensor data, the framework segments execution into approach, alignment, and insertion stages for a Peg in Hole (PiH) assembly operation. By using goal relative position data as features, this classification generalizes to unseen goal positions without requiring retraining, matching the inherent generalization capability of DMPs. Anomaly detection is performed online using a Mahalanobis distance metric computed over force features, isolating contact induced failures from nominal trajectory execution. Upon failure detection, a spiral search recovery policy is triggered to actively realign the peg under contact before resuming the learned DMP insertion. The proposed approach is evaluated on an experimental setup achieving 95\% stage classification accuracy, and demonstrates reliable failure recovery under lateral misalignments of up to 3~mm using only a single demonstration.

\end{abstract}

\section{INTRODUCTION}
Robotic manipulation plays a crucial role in industrial automation, particularly in tasks that require precise motion, adaptability, and physical interaction with the environment~\cite{INTRO}. However, programming robots to operate with precision and adaptability simultaneously can be challenging. Existing industrial automation widely relies on manual programming, with specification of exact waypoints which can be time consuming, lacks adaptability, and fails to generalize for even small changes in task settings \cite{INTRO2}.

This motivates the use of Learning from Demonstration (LfD), also referred to as Imitation Learning (IL), where the robot learns the desired behaviour by imitating an expert~\cite{LFD3}. The choice of LfD is particularly compelling when the desired robot behavior cannot be easily scripted through conventional programming but can be naturally demonstrated by a human expert~\cite{LFD_BILLARD_1}. By observing such demonstrations, the robot can learn task knowledge, motion patterns, and manipulation strategies, and reproduce similar trajectories under comparable but slightly varying conditions~\cite{LFD2}.


\begin{figure*}[!t]
    \centering
    \includegraphics[width=1.8\columnwidth]{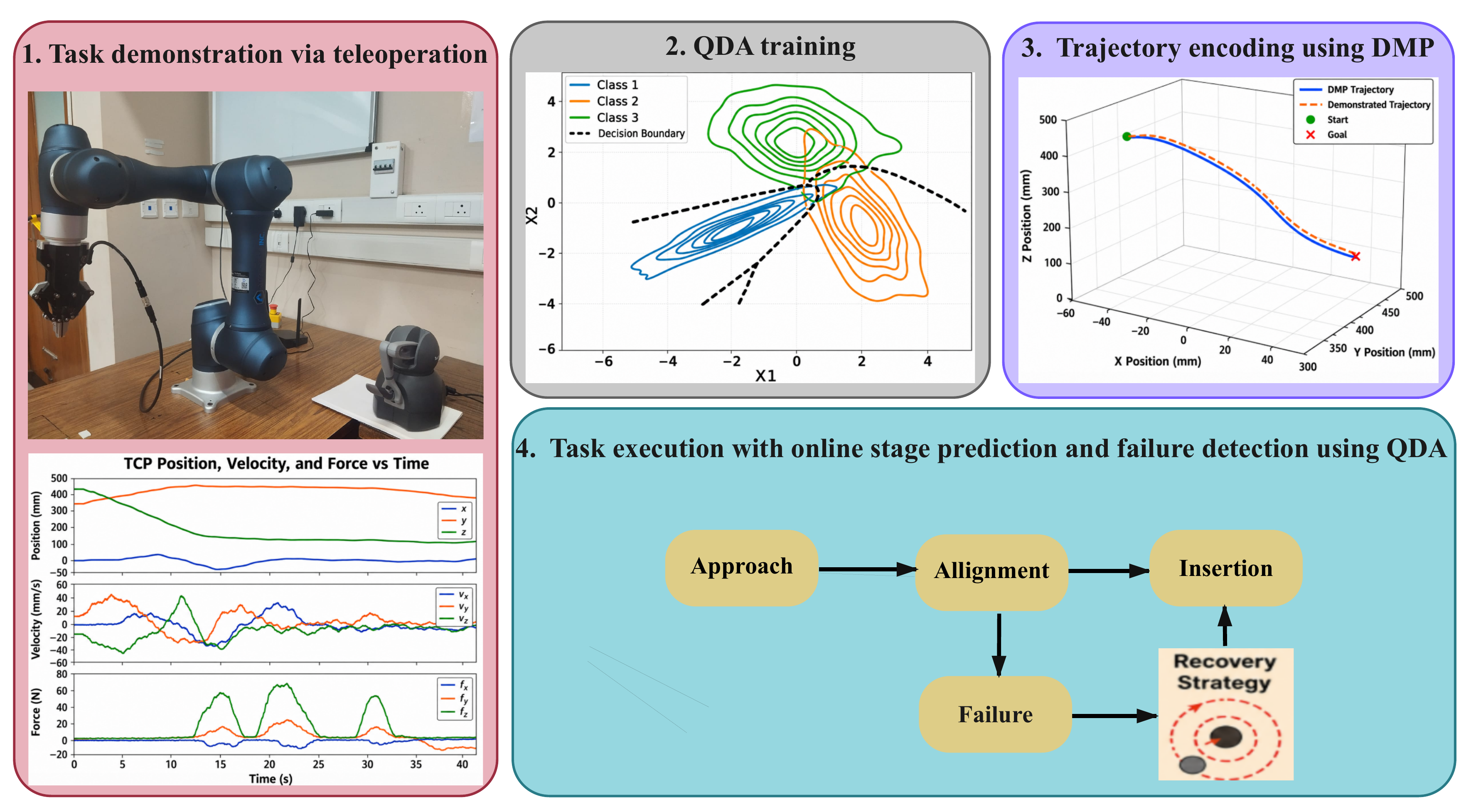}
    \caption{Overview of proposed learning and recovery framework, starting with 1. single task demonstration through teleoperation, 2. QDA training for stage classification, 3. encoding the trajectory using dynamic movement primitive, and 4. the proposed framework for the Peg in Hole (PiH) task. }
    \label{fig:flowchart}
\end{figure*}
Within the paradigm of LfD, Dynamic Movement Primitives (DMPs) have been widely used to encode and reproduce demonstrated robot motions~\cite{dmp}. DMPs represent the demonstrated trajectory as a stable second order dynamical system, allowing the robot to preserve the overall shape of the learned motion while adapting to changes in start position, goal position, and execution speed~\cite{DMP2}. However, standard DMP formulations are often limited when task conditions vary, such as changes in object pose, contact location, or task geometry \cite{limitations}. These variations can reduce reproduction accuracy and may lead to task failure, particularly in contact rich manipulation. In such cases, stage segmentation provides useful task level awareness by dividing the execution into meaningful phases. By identifying the current stage, the robot can better interpret the execution context, detect stage specific deviations, and apply appropriate recovery actions rather than relying only on trajectory encoding using DMPs. This limitation motivates the integration of stage classification into DMP based manipulation frameworks.

Stage segmentation has been explored in robotic assembly to capture the sequential structure of task execution. In~\cite{HMM}, a Hidden Markov Model (HMM) framework is employed for assembly stage estimation, where task progression is modeled through transitions between discrete stages. However, online stage inference can be challenging near stage transitions, as future observations are unavailable and adjacent stages may exhibit overlapping force position characteristics. Moreover, their formulation relies on residual equations to define individual stages, which is feasible in constrained robotic assembly but difficult to generalize to human demonstrations due to natural variability in execution \cite{billard2}, \cite{HMM2}. Previous works have also demonstrated that  different phases of assembly exhibit distinct temporal signatures, and that learning based methods can effectively map sensor trajectories to success or failure outcomes \cite{Aronson2016DataDrivenCO}. Together, these studies underscore the relevance of data driven stage classification for contact rich manipulation and motivate methods that can infer task phases directly from observed multimodal sensor data.

The current article introduces a stage aware framework for failure detection and recovery that integrates DMP based trajectory generation with real time stage classification and force based anomaly detection. Quadratic Discriminant Analysis (QDA)~\cite{QDA} classifies execution stages using multimodal sensor inputs, enabling structured interpretation of task progression. Failure detection is performed using a statistical distance measure computed over force features for reliable identification of contact induced anomalies. Based on the detected stage and failure condition, a stage specific recovery action is performed to improve robustness during task execution.

This capability is particularly important for Peg in Hole (PiH) insertion, a widely used benchmark task in robotic assembly~\cite{peginhole, sangwoon2022icra, yasutomi2021icra}. Despite its geometric simplicity, the task remains challenging due to complex contact rich interactions and sensitivity to small variations in initial alignment and grasp configuration. Classical studies on rigid part mating have shown that assembly success is strongly influenced by part geometry, and initial lateral or angular misalignment, which can lead to contact induced phenomena such as jamming and wedging during insertion~\cite{10.1115/1.3149634}.

PiH insertion task involves distinct stages such as approach, alignment, and insertion. These stages exhibit different position and force characteristics, and without explicit stage modeling, it becomes difficult to interpret execution context and to identify task failures in real time. Thus, by combining motion reproduction, stage recognition, and force based recovery, the proposed framework provides a robust foundation for PiH insertion and similar contact rich manipulation tasks that require adaptive and context aware execution.

The main contributions of this work include:
\begin{enumerate}[]
    \item A QDA based real time stage classification method using multimodal sensor data for a benchmark contact rich task i.e. PiH insertion.
    \item A data efficient learning framework based on DMPs trained from a single demonstration with failure detection and recovery.
\end{enumerate}

The remainder of this paper is organized as follows: Section \ref{sec:methodology} describes the detailed methodology with mathematical background, Section \ref{sec:experiment} describes the hardware setup used for performing the contact rich manipulation task and Section \ref{sec:results} discusses the results, followed by conclusion in Section \ref{sec:conclusion}. 

\section{Methodology}\label{sec:methodology}

\subsection{Task Demonstration}
The robot is guided via teleoperation to demonstrate the PiH insertion task while recording End Effector (EE) position, velocity, and force data at a sampling frequency of 1000 Hz. These demonstrated cartesian trajectories serve as reference for trajectory tracking, stage classification, and anomaly detection during task execution. Since, the focus of this work is on failure detection and recovery during execution rather than trajectory variability, a single demonstration is used as the reference trajectory. An overview of the proposed learning and recovery framework is given in Figure~\ref{fig:flowchart}.

\subsection{Task Segmentation}
PiH assembly is a complex contact interaction task which cannot be captured by a single continuous trajectory. Therefore, we segment the process is into three semantically meaningful stages: \textit{Approach}, \textit{Alignment}, and \textit{Insertion}.


\begin{enumerate}
    \item \textbf{Approach:} The robot moves the peg towards the hole from a home position.
    \item \textbf{Alignment:} The peg makes the initial contact with the hole and adjusts its position to align with the hole axis.
    \item \textbf{Insertion:} The peg is inserted into the hole along the aligned direction.
\end{enumerate}

Due to variations in the initial grasp position or environmental uncertainties, failure may occur in the \textit{Alignment} stage during execution of the learnt DMP trajectory.

\begin{figure}[t]
    \centering
    \includegraphics[width=0.3\textwidth]{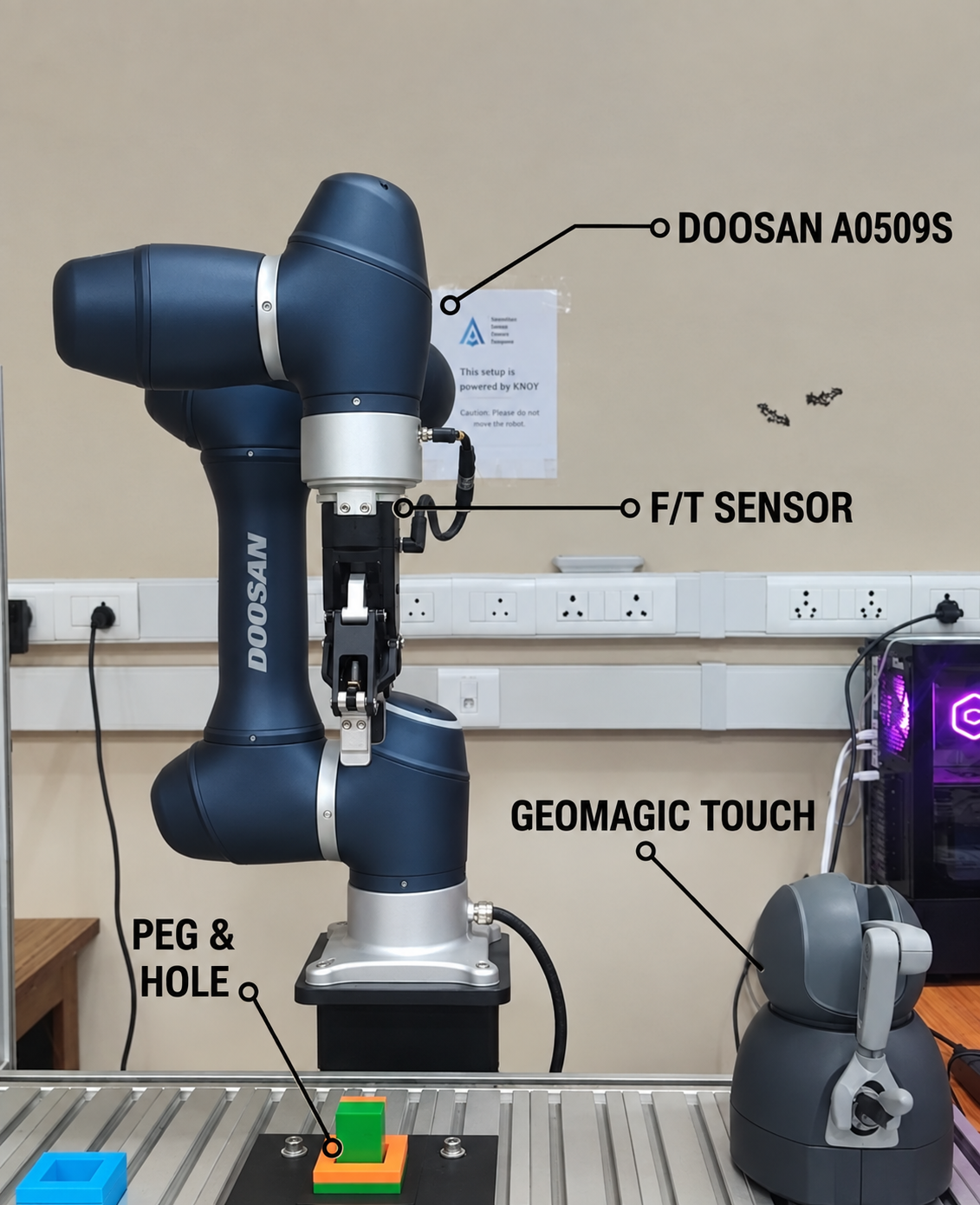}
    \caption{Experimental setup for collecting task demonstrations via teleoperation}
    \label{fig:teleop}
\end{figure}
\subsection{Dynamic Movement Primitives (DMPs)}
DMPs are used to encode and learn the demonstrated trajectories using a simple second order spring, mass, damper model driven by a non linear forcing term. For the PiH insertion task, a separate DMP is learnt for each spatial dimension $(x,y,z)$ using the recorded demonstration trajectory.

For a single Degree of Freedom (DoF), it is defined as:
\begin{equation}
\tau^2 \ddot{x}_i =
\alpha_x \big( \beta_x (g_i - x_i) - \tau \dot{x}_i \big)
+ f_i(s)
\end{equation}
\noindent
where $x_i$ denotes the position along dimension $i \in \{x,y,z\}$, $g_i$ is the goal position, and $\dot{x}_i$ and $\ddot{x}_i$ represent velocity and acceleration, respectively, and $f_i(s)$ is the non linear forcing function. The parameters $\alpha_x$ and $\beta_x$ define the spring, and damper behavior of the system, $\tau$ is a temporal scaling factor controlling execution speed. The forcing term $f_i(s)$ is represented as a weighted sum of Gaussian basis functions,

\begin{equation}
    f_i(s) = \frac{\sum_{i=1}^{N} w_i \psi_i(s)}{\sum_{i=1}^{N} \psi_i(s)} s
\end{equation}
\noindent
where $\psi_i(s) = \exp(-h_i (s - c_i)^2)$ are Gaussian basis functions with centers $c_i$ and widths $h_i$, and $w_i$ are learnable weights.

To achieve time independent trajectory generation, DMPs employ a canonical system governed by a phase variable $s$:
\begin{equation}
\tau \dot{s} = -\alpha_s s
\end{equation}
\noindent
where $\alpha_s$ is a positive decay constant. The phase variable monotonically decreases from $1$ to $0$, synchronizing the evolution of all trajectory dimensions during execution.

The learned DMPs are executed online to generate the reference motion for PiH insertion.

\subsection{ Quadratic Discriminant Analysis (QDA)}

Stage classification is formulated as a supervised learning problem using QDA, trained on the following features extracted from the demonstrated trajectory,
\begin{itemize}
    \item relative EE position $(x - x_g, \; y - y_g, \; z - z_g)$,
    \item force measurements $(F_x, F_y, F_z)$, and,
    \item EE velocity $(v_x, v_y, v_z)$.
\end{itemize}

\noindent

The relative position is computed with respect to the goal position and is consistent with the goal relative representation used in DMPs. 
Each data sample is represented by a feature vector $u$, which contains the data used for stage classification formed by concatenating the relative position with respect to the goal, the force measurements, and the EE velocity,
\begin{equation}
u =
\begin{bmatrix}
(p-p_g)^{T} &
F^{T} &
v^{T}
\end{bmatrix}^{T},
\end{equation}
where
\begin{align}
&p=[x,y,z]^T,
\qquad
&p_g=[x_g,y_g,z_g]^T, \\
&F=[F_x,F_y,F_z]^T,
\qquad
&v=[v_x,v_y,v_z]^T.
\end{align}

The feature vector corresponding to stage $k$ is assumed to follow a multivariate Gaussian distribution:
\begin{equation}
p(u|k) \sim \mathcal{N}(\mu_k,\Sigma_k),
\end{equation}
where $\mu_k$ denotes the mean feature vector of stage $k$, representing the average value of each feature across the training samples belonging to that stage, and $\Sigma_k$ denotes the covariance matrix of stage $k$, which characterizes the variability of the features and their pairwise correlations. The parameters $\mu_k$ and $\Sigma_k$ are estimated from the demonstrated trajectories associated with stage $k$.

The stage label is obtained as:
\begin{equation}
L = \arg\max_k \; p(k|u)
\end{equation}

where, $L \in \{\text{Approach}, \text{Alignment}, \text{Insertion}\}$. The predicted stage is used to enable stage aware monitoring and recovery during execution.
\begin{figure}[t]
    \centering
    \includegraphics[width=0.3\textwidth]{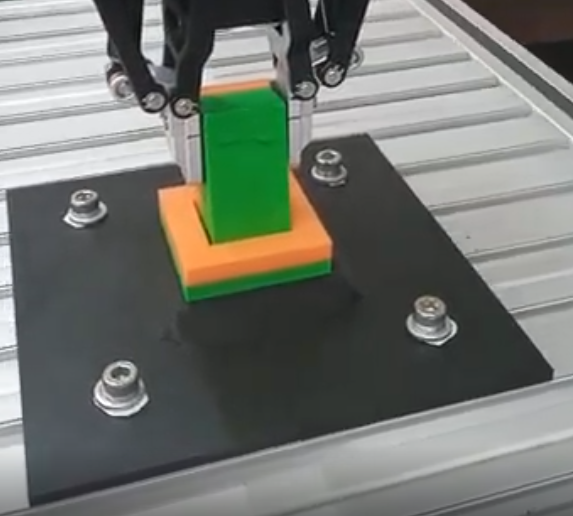}
    \caption{3D-printed peg in hole insertion setup for demonstrating contact rich task}
    \label{fig:peg_setup}
\end{figure}
\subsubsection{Failure Detection}
Failures are detected using a Mahalanobis distance ~\cite{mahalanobis} based criterion computed in the force feature space. Let $\mathbf{u}_F = [F_x, F_y, F_z]^T$ denote the contact force vector. For the predicted stage $k$, the distance is defined as:
\begin{equation}
d(\mathbf{u}_F) = (\mathbf{u}_F - \mu_k)^T \Sigma_k^{-1} (\mathbf{u}_F - \mu_k)
\end{equation}

A failure is declared when:
\begin{equation}
d(\mathbf{u}_F) > \mu_d + 6\sigma_d
\end{equation}

Here, $\mu_d$ and $\sigma_d$ are the mean and standard deviation of Mahalanobis distances estimated from recorded demonstration data. The six sigma factor was selected empirically ensuring that normal contact transitions are not misclassified as failures while still detecting significant deviations due to misalignment. In practice, this threshold provided the most stable performance across multiple trials without requiring manual tuning per stage.

\vspace{0.5em}
\noindent
\subsubsection{Feature Selection for Failure Detection}
While all features are used for stage classification, failure detection is performed using only force features $(F_x, F_y, F_z)$. The position trajectory is accurately tracked by the DMP even during failure. 

As a result, errors in peg pose or grasp configuration are not reflected in the position signal but instead manifest as abnormal contact forces.  
Therefore, restricting failure detection to force features gives a more physically meaningful criterion.
\begin{figure}[t]
\centering
\includegraphics[width=\linewidth]{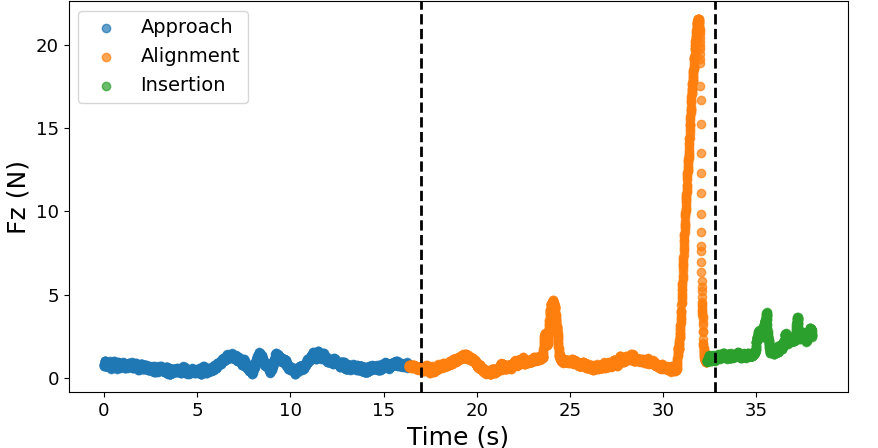}
\caption{Stage prediction using QDA on DMP execution}
\label{fig:qda_prediction}
\end{figure}
\subsection{Recovery Action}

Failures primarily occur due to the lateral misalignment of peg with reference to original grasping pose, during the \textit{Alignment} stage. Upon failure detection using the Mahalanobis distance criterion, the DMP execution is paused and a recovery policy is triggered.

The recovery strategy performs a spiral search in the $(x,y)$ plane under contact to relocate the hole. The spiral motion is defined in cartesian coordinates as,

\begin{equation}
x(t) = kt\cos(\omega t), \quad y(t) = kt\sin(\omega t),
\end{equation}
\noindent
where $k$ controls radial expansion and $\omega$ defines angular speed. The search radius is bounded to ensure stable interaction.

Alignment is detected using force feedback. Specifically, a reduction in lateral contact forces $(F_x, F_y)$ together with a decrease in axial force along the insertion direction $(F_z)$ indicates successful alignment. The system infers alignment when the axial force stabilizes near zero and lateral forces remain consistently low.

Once alignment is detected, recovery is terminated and the robot resumes the DMP learnt trajectory.

\section{Experimental Setup} \label{sec:experiment}
Human demonstrations are collected through teleoperation using the setup shown in Figure~\ref{fig:teleop}. The experimental platform consists of a Doosan A0509s robotic arm with a force-torque sensor at the flange, a Geomagic Touch haptic device, and a two finger DH-AG95 gripper. The Geomagic Touch acts as a master device during task demonstration, while the robotic manipulator constitutes the slave system. The teleopsystem operates in velocity control mode, mapping cartesian velocity commands of the master to the slave.

The complete experimental framework is implemented in ROS~2 Humble on Ubuntu~22.04. ROS~2 serves as the underlying software and communication framework for the entire system, integrating the haptic interface, robot control, task demonstration, and autonomous task execution. During each demonstration, the cartesian position and velocity of the robot EE, together with the measured interaction forces, are recorded at a sampling frequency of 1000~Hz and stored as a ROS~2 bag file for offline processing. 


The experiments are conducted using a 3D-printed PiH setup as shown in Figure~\ref{fig:peg_setup}. A square peg of size $33 \times 33~\text{mm}$ is inserted into a $35 \times 35~\text{mm}$ hole, providing $1~\text{mm}$ clearance per side and requiring precise alignment under contact.

The peg is grasped using a two finger parallel gripper, while the hole fixture is rigidly mounted to ensure a fixed reference frame. The small clearance between the peg and the hole require accurate translational, and rotational allignment during insertion, providing a contact rich manipulation setting for evaluating the proposed framework.

\section{Results}\label{sec:results}

The proposed stage aware framework is evaluated on teleoperated demonstrations and DMP executed trajectories.

\subsubsection{Demonstration and Stage Learning}

\begin{figure}[t]
\centering
\includegraphics[width=\linewidth]{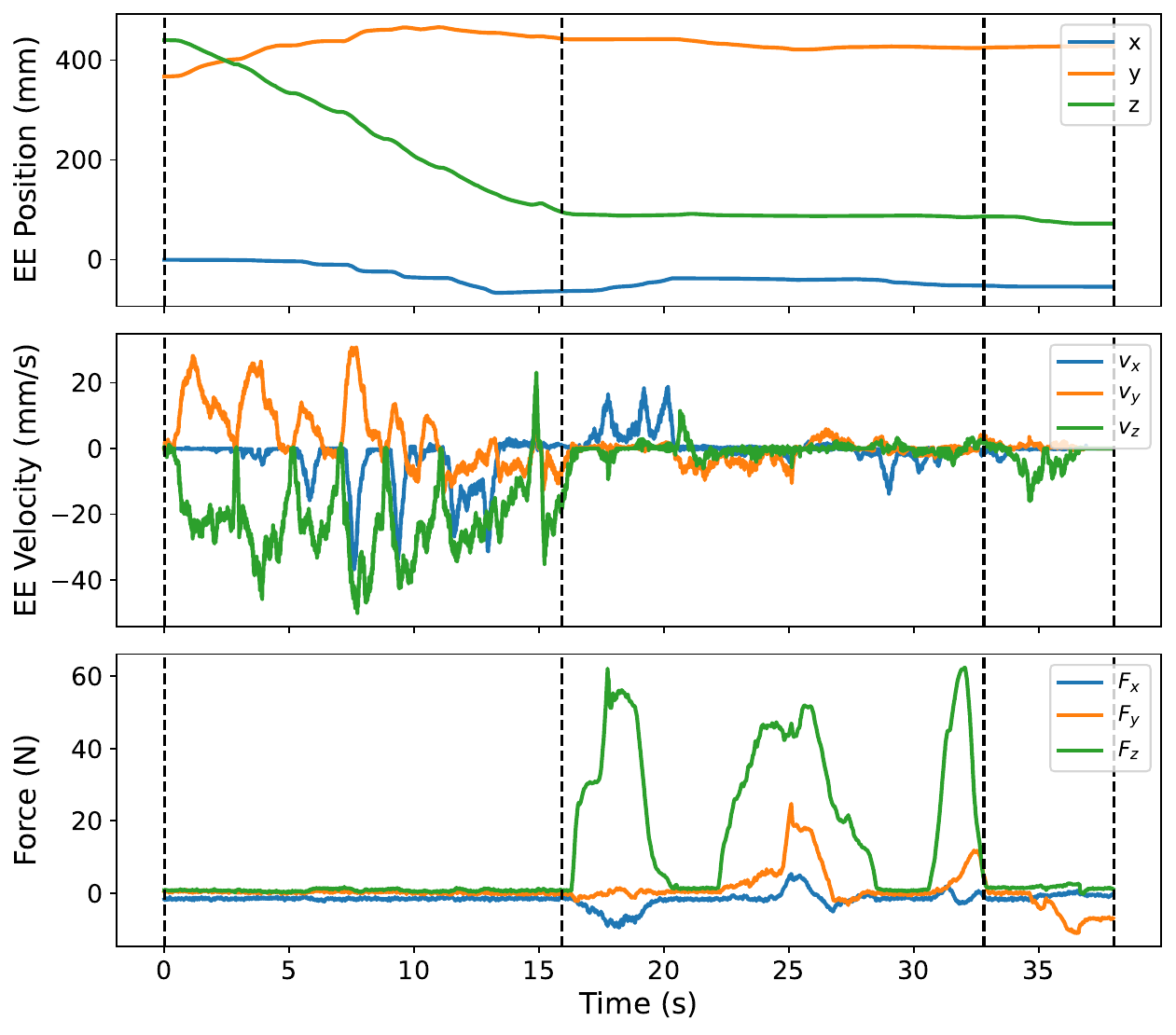}
\caption{Demonstration data with time labeled task stages used for training}
\label{fig:demo_labels}
\end{figure}

Figure~\ref{fig:demo_labels} shows the recorded demonstration with manually labeled task stages, namely \textit{Approach}, \textit{Alignment}, and \textit{Insertion}.  These labeled segments are used to train both the DMPs for trajectory reproduction and the QDA classifier for stage recognition.

\subsubsection{Stage Classification}

The performance of the QDA based stage classifier is shown in Figure~\ref{fig:qda_prediction}. 
The predicted stage labels closely follow the demonstrated segmentation, indicating that the classifier reliably distinguishes between different phases of the task during execution. 
The QDA based stage classifier was ran over multiple trials as dicussed in Table~\ref{tab:qda_trials} and achieves an overall accuracy of 95\% during execution, as detailed in Table~\ref{tab:time_comparison}.
\begin{table}[!t]
\caption{Stage Classification Accuracy Across Recorded Execution Trials}
\centering
\begin{tabular}{|l|c|}
\hline
\textbf{Trial} & \textbf{Accuracy} (\%) \\
\hline
1 & 97.5 \\
2 & 88.7 \\
3 & 95.0 \\
4 & 86.1 \\
5 & 85.4 \\
6 & 98.0 \\
\hline
Mean $\pm$ Std & $91.8 \pm 5.4$ \\
\hline
\end{tabular}
\label{tab:qda_trials}
\end{table}

\begin{table}[!t]
\caption{Comparison of True and Predicted Stage Time Intervals}
\begin{center}
\begin{tabular}{|l|c|c|}
\hline
Stage & True Time Interval (s) & Predicted Time Interval (s) \\
\hline
Approach    & $0.0$ -- $17.00$ & $0.0$ -- $16.36$ \\
Alignment   & $17.00$ -- $32.80$ & $16.36$ -- $32.34$ \\
Insertion   & $32.80$ -- $38.00$ & $32.34$ -- $38.00$ \\
\hline
Overall Accuracy & \multicolumn{2}{|c|}{95\%} \\
\hline
\end{tabular}
\end{center}
\label{tab:time_comparison}
\end{table}

\begin{figure}[!ht]
\centering
\includegraphics[width=\linewidth]{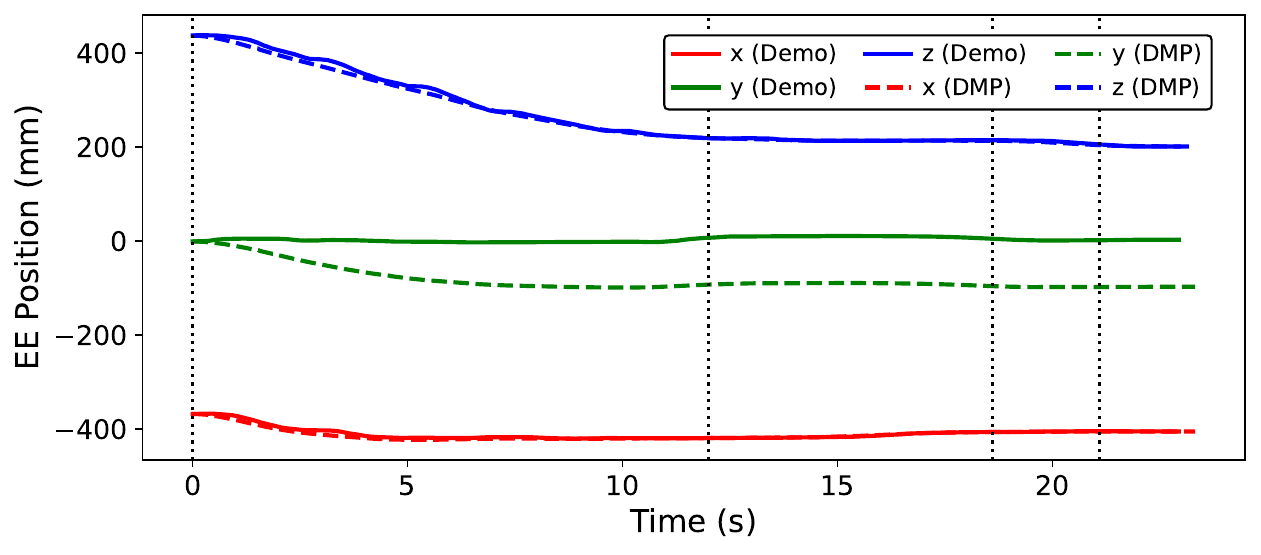}
\caption{Comparison of demonstration trajectory and DMP execution under a shifted goal position}
\label{fig:goal_shift_traj}
\end{figure}

The variation in classification accuracy is mainly observed near stage transitions, where force-position characteristics overlap due to contact interaction.

To evaluate generalisation to different goal, the classifier is tested for a modified goal position, where the original goal $(-400, 0, 200)$ is shifted to $(-400, -100, 200)$.

Figure~\ref{fig:goal_shift_traj} shows the comparison between the original demonstration trajectory and the DMP executed trajectory under the shifted goal. 
\begin{figure}[t]
\centering
\includegraphics[width=\linewidth]{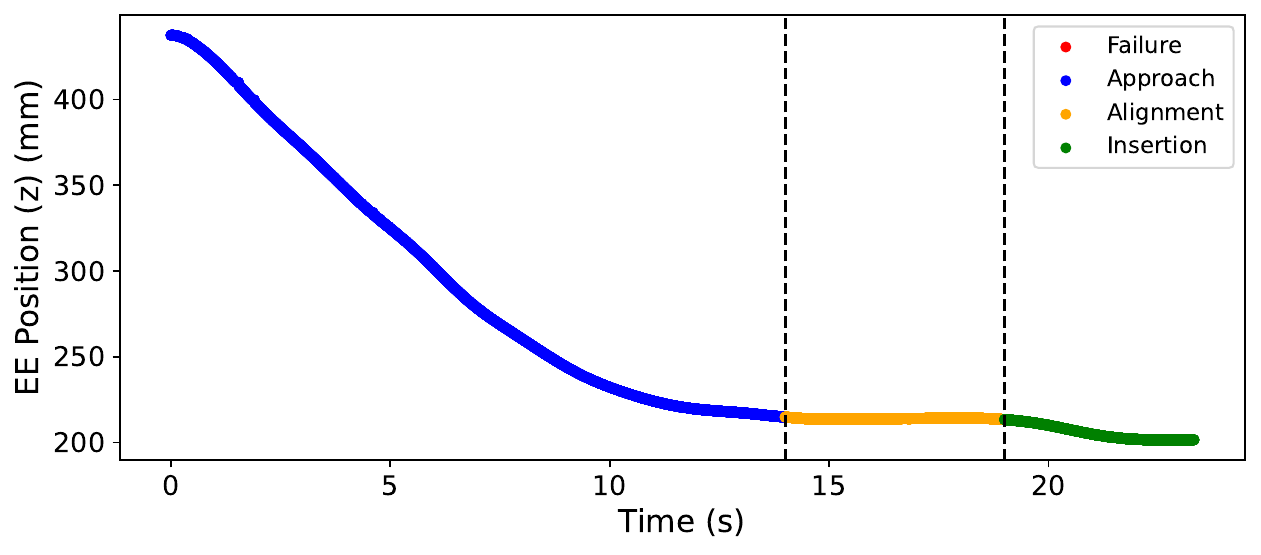}
\caption{Stage classification results on DMP execution with shifted goal position where the classifier correctly generalises across goal variations.}
\label{fig:goal_shift_classification}
\end{figure}

\begin{figure}[!ht]
\centering
\includegraphics[width=\linewidth]{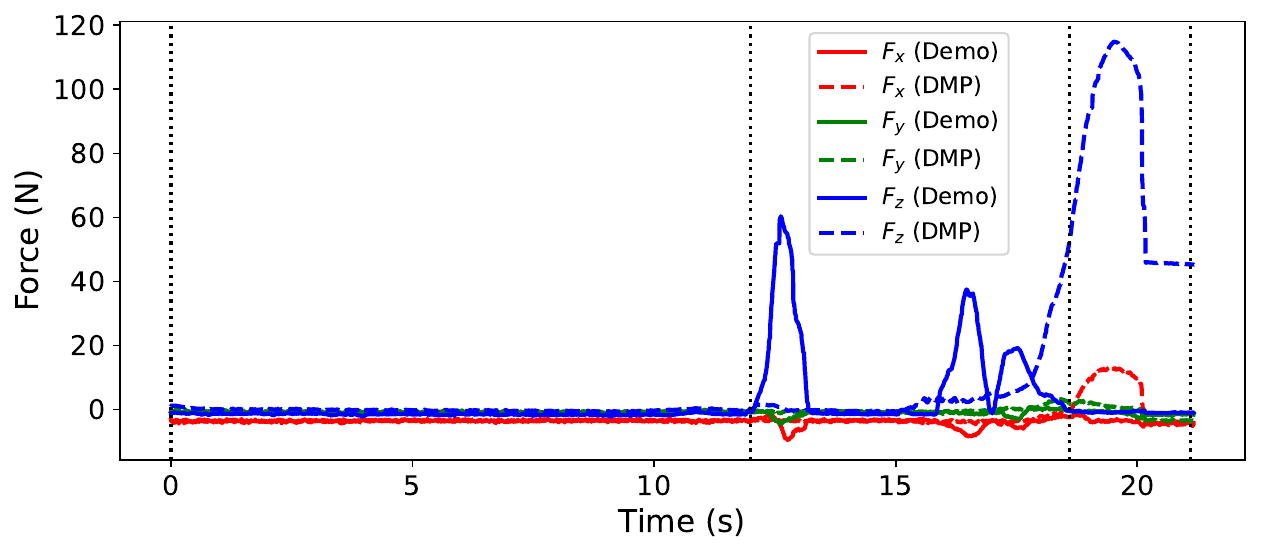}
\caption{Comparison of force profiles between demonstration and execution.}
\label{fig:demo_comparison}
\end{figure}

As shown in Figure~\ref{fig:goal_shift_classification}, the classifier maintains consistent stage segmentation under the shifted goal without retraining, correctly identifying \textit{Approach}, \textit{Alignment}, and \textit{Insertion} phases. 
This confirms that goal relative position feature enables generalisation across goal variations.
\subsubsection{Failure Detection}

The demonstration corresponds to a failure case occurring during the allignment stage, which leads to excessive force buildup along the insertion axis ($F_z$). The robot execution is compared against the demonstrated trajectory in terms of position, velocity, and force profiles, as shown in Figure~\ref{fig:demo_comparison}. The deviation in force response, despite similar nominal motion trends, highlights the onset of failure during contact rich interaction.

\begin{figure}[!ht]
\centering
\includegraphics[width=\linewidth]{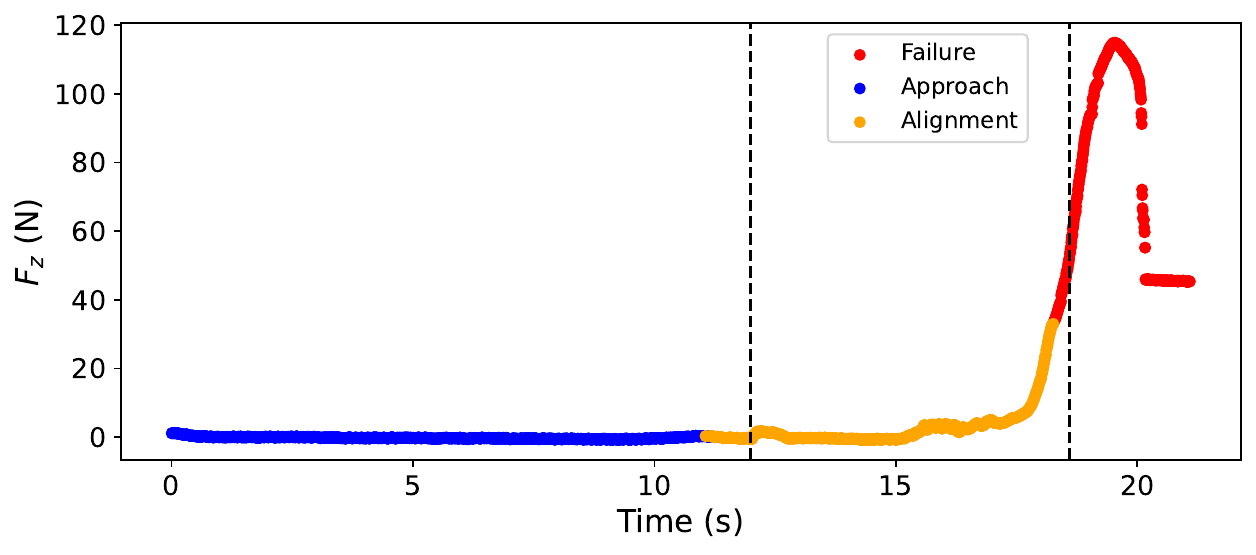}
\caption{Sharp rise in force along z direction during insertion stage due to failure.}
\label{fig:force_detection}
\end{figure}

Figure~\ref{fig:force_detection} shows the force profile along the insertion axis ($F_z$) over time. A sharp increase in contact force is observed during the insertion phase, which corresponds to the failure event. The proposed detection mechanism identifies this region based on force deviation from nominal behavior.

\begin{figure}[!ht]
\centering 
\includegraphics[width=\linewidth]{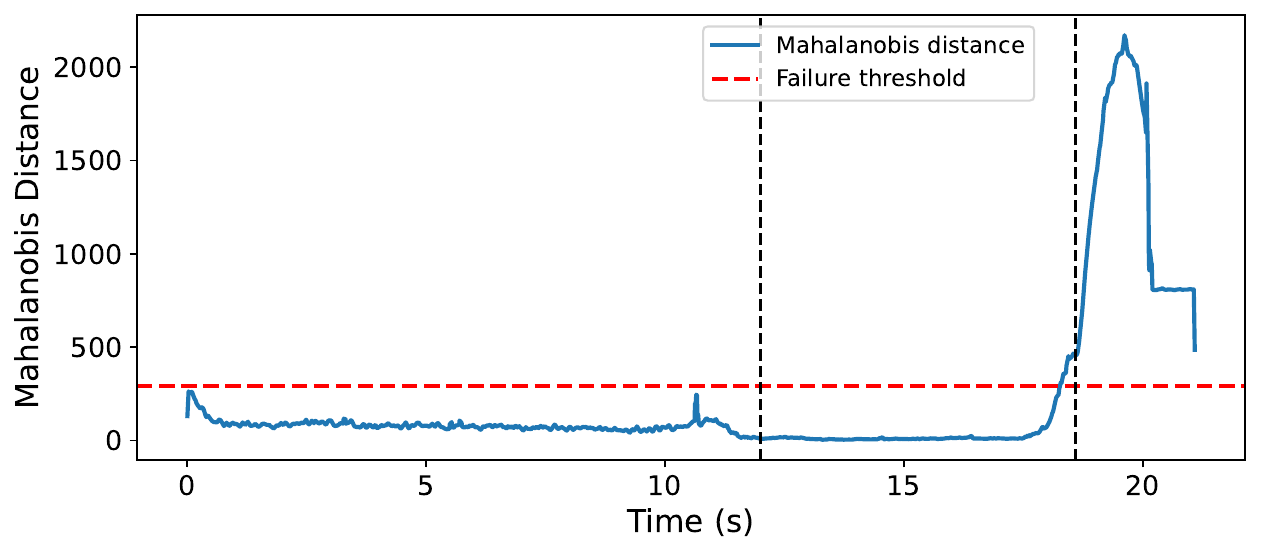}
\caption{Mahalanobis distance during execution with six-sigma threshold.}
\label{fig:mahalanobis}
\end{figure}

To formally validate the detection approach, the Mahalanobis distance is computed over the feature space, as shown in Figure~\ref{fig:mahalanobis}. A significant rise in distance is observed during the failure phase, exceeding the six-sigma threshold and enabling reliable detection of the failure condition.

\subsection{Recovery Action}

To evaluate robustness, recovery experiments are conducted under induced misalignment, where the peg is initially offset by approximately $3~\text{mm}$ from the hole center, causing failure during the alignment stage.

Upon failure detection, a spiral search is executed in the $(x,y)$ plane to reestablish alignment. The trajectory is defined in Cartesian coordinates as:

\begin{equation}
x(t) = kt\cos(\omega t), \quad y(t) = kt\sin(\omega t)
\end{equation}

The parameters are set as $k = 0.0002~\text{m/s}$, $\omega = 2.0~\text{rad/s}$, and maximum spiral radius equal to $8~\text{mm}$ to ensure local and stable exploration.

Alignment is inferred from force feedback, where reduction in lateral forces and stabilization of axial force $F_z$ indicate successful centering of the peg within the hole.

The proposed recovery strategy successfully restores insertion under up to $3~\text{mm}$ misalignment, demonstrating improved robustness in contact-rich assembly tasks.

\section{Conclusion and Future Work}
\label{sec:conclusion}

In this work, we presented a stage aware framework that integrates DMPs with QDA for real time stage classification and failure detection of PiH insertion task. The system is trained on a single teleoperated human demonstration, decomposed into \textit{approach}, \textit{alignment}, and \textit{insertion} stages. DMPs track the position trajectory, QDA is used for online stage classification, while a Mahalanobis distance based criterion is used to detect force anomalies associated with failure conditions.

To handle failures during \textit{alignment} stage, a spiral search recovery strategy is introduced that allows the robot to realign and then resume the learned DMP trajectory. Experimental results on a real robotic setup show successful PiH insertion task execution, failure detection and recovery up to 3~mm offset from hole center, demonstrating the effectiveness of the proposed framework for robust precision assembly.

Future work will focus on integrating vision feedback and extending the framework to handle more complex failure modes and tighter tolerance assembly tasks.

\addtolength{\textheight}{-12cm}   





\section*{ACKNOWLEDGMENT}
The authors would like to acknowledge support in part by Anusandhan National Research Foundation (ANRF) under grant SP25260420MEANRF009063.


\end{document}